\documentclass[times, 10pt]{article}
\usepackage{latex8}
\usepackage{times}
\usepackage[authoryear]{natbib}
\usepackage{graphicx}
\usepackage{amsfonts}
\usepackage{subfigure}
\usepackage{setspace}
\usepackage{color}
\usepackage{verbatim}
\usepackage{algorithm}
\usepackage{algorithmic}
\usepackage{multirow}
\usepackage{latexsym}
\usepackage{wrapfig}

\newtheorem{definition}{Definition}

\renewcommand{\cite}{\citep}
\begin{document}

\title{SparseDTW: A Novel Approach to Speed up Dynamic Time Warping}

\author{Ghazi Al-Naymat$^{1 , }$\thanks{The work was done while the author at School of Information Technologies, The University of Sydney, Australia.}
\and 
Sanjay Chawla$^{2}$
\and Javid Taheri$^{2}$}
\affiliation{$^1$ School of Computer Science and Engineering \\
The University of New South Wales\\
Sydney, NSW 2052, Australia \\
Email:~{\tt ghazi@cse.unsw.edu.au}\\[.1in]
$^2$ School of Information Technologies \\
The University of Sydney, Australia\\
Email:~{\tt \{chawla,javidt\}@it.usyd.edu.au}}

\maketitle




\begin{abstract}
We present a new space-efficient approach, (\emph{SparseDTW}), to
compute the Dynamic Time Warping (\emph{DTW}) distance between two
time series that always yields the optimal result. This is in
contrast to other known approaches which typically sacrifice
optimality to attain space efficiency. The main idea behind our
approach is to dynamically exploit the  existence of similarity
and/or correlation between the time series. The more the similarity
between the time series the less space required to compute the
\emph{DTW} between them. To the best of our knowledge, all other
techniques to speedup \emph{DTW}, impose apriori constraints and do
not exploit similarity characteristics that may be present in the
data. We conduct experiments and demonstrate that \emph{SparseDTW}
outperforms previous approaches.  
\end{abstract}
\vspace{.1in}


\section{Introduction} \label{Sec:TSM-Introduction}


Dynamic time warping (\emph{DTW}) uses the dynamic programming
paradigm to compute the alignment between two time series. An {\it
alignment} ``warps'' one time series onto another and can be used as
a basis to determine the similarity between the time series.
\emph{DTW} has similarities to sequence alignment in bioinformatics
and computational linguistics except that the {\it matching} process
in sequence alignment and {\it warping} have to satisfy a different
set of constraints and there is no gap condition in warping.
\emph{DTW} first became popular in the speech recognition
community~\citep{DTW-Originator} where it has been used to determine
if the two speech wave-forms represent the same underlying spoken
phrase. Since then it has been adopted in many other diverse areas
and has become the similarity metric of choice in time series
analysis~\citep{EamonnScalingupDTW}.

Like in sequence alignment, the standard \emph{DTW} algorithm has
$O(mn)$ space complexity where $m$ and $n$ are the lengths of the
two sequences being aligned. This limits the practicality of the
algorithm in todays ``data rich environment'' where long sequences
are often the norm rather than the exception. For example, consider
two time series which represent stock prices at one second
granularity. A typical stock is traded for at least eight hours on
the stock exchange and that corresponds to a length of $8 \times 60
\times 60 = 28800$. To compute the similarity, \emph{DTW} would have
to store a matrix with at least 800 million entries!

Figure~\ref{Fig:TSM-WarpingA} shows an example of an alignment
(warping) between two sequences $S$ and $Q$. It is clear that there
are several possible alignments but the challenge is to select the
one which has the minimal overall distance. The alignment has to
satisfy several constraints which we will elaborate on in
Section~\ref{Sec:TSM-DTWBackground}.

~\citet{FastDTW04} have provided a succinct categorization of
different techniques that have been used to speed up \emph{DTW}:

\begin{itemize}
  \item \textbf{Constraints}: By adding additional constraints the search
    space of possible alignments can be reduced. Two well known
    exemplars of this approach are the~\citet{DTW-Originator} and the~\citet{Itakura75} constraints which limit
    how far the alignment can deviate from the diagonal. While these
    approaches provide a relief in the space complexity, they do not
    guarantee the optimality of the alignment.

  \item \textbf{Data Abstraction}: In this approach, the warping path is
    computed at a lower resolution of the data and then mapped back to
    the original resolution \citep{FastDTW04}. Again, optimality of the
    alignment is not guaranteed.

  \item \textbf{Indexing}:~\citet{EamonnIdxDTW05},~\citet{FTW05}, and~\citet{Lemire-FastDTW} proposed an
    indexing approach, which does not directly speed up \emph{DTW} but
    limits the number of \emph{DTW} computations. For example, suppose
    there exists a database D of time series sequences and a query
    sequence $q$. We want to retrieve all sequences $d \in D$ such
    that $DTW(q,d) < \epsilon$. Then instead of checking $q$ against
    each and every sequence in $D$, an easy to calculate lower bound
    function LBF is first applied between $q$ and $D$. The argument
    works as follows:
    \begin{enumerate}
        \item By construction, $LBF(q,d) < DTW(q,d)$.
        \item Therefore, if $LBF(q,d) > \epsilon$ then $DTW(q,d) > \epsilon$ and $DTW(q,d)$ does not have to be computed.
    \end{enumerate}
\end{itemize}

\subsection{Main Contribution}

The main insight behind our proposed approach, \emph{SparseDTW}, is
to dynamically exploit the possible existence of inherent similarity
and correlation between the two time series whose \emph{DTW} is
being computed. This is the motivation behind the Sakoe-Chiba band
and the Itakura Parellelogram
but our approach has three distinct advantages:\\
\begin{enumerate}
    \item Bands in \emph{SparseDTW} evolve dynamically and
        are, on average, much smaller than the traditional approaches.
        We always represent the warping matrix using  sparse
        matrices, which leads to better average space complexity  compared to other approaches (Figure~\ref{Fig:TSM-AllDTWAlgos}).
    \item \emph{SparseDTW} always yields the optimal warping
        path since we never have to set apriori constraints independently of
        the data. For example, in the traditional banded approaches, a
        sub-optimal path will result if all the possible optimal warping
        paths have to cross
        the bands.
    \item Since \emph{SparseDTW} yields an optimal alignment, it
        can easily be used in conjunction with lower bound approaches.
\end{enumerate}

\subsection{Paper Outline}

The rest of the paper is organized as follows: Section
\ref{Sec:TSM-RelatedWork} describes related work on \textit{DTW}.
The \textit{DTW} algorithm is described in Section
\ref{Sec:TSM-DTWBackground}. In Section~\ref{Sec:TSM-BandDTW}, we
give an overview of the techniques used to speed up DTW by adding
constraints. Section~\ref{Sec:TSM-DC} reviews the Divide and Conquer
approach for \emph{DTW} which is guaranteed to take up $O(m+n)$
space and $O(mn)$ time. Furthermore, we provide an example which
clearly shows that the divide and conquer approach fails to arrive
at the optimal \emph{DTW} result. The \emph{SparseDTW} algorithm is
introduced with a detailed example in
Section~\ref{Sec:TSM-SparseDTW}. We
analyze and discuss our results in Section~\ref{Sec:TSM-Exps}, followed by our
conclusions in Section~\ref{Sec:PT-Summary}.



\section{Related Work} \label{Sec:TSM-RelatedWork}

DTW was first introduced in the data mining community in the context
of mining time series \citep{Berndt94}. Since it is a flexible
measure for time series similarity it is used extensively for ECGs
(Electrocardiograms) \citep{ECG98-journal}, speech processing
\citep{Rabiner-Book}, and robotics \citep{schmill99learned}. It is
important to know that \emph{DTW} is a measure not a metric, because
\emph{DTW} does not satisfy the triangular inequality.

Several techniques have been introduced to speed up \emph{DTW}
and/or reduce the space overhead
\citep{Hirschberg75,LB-Yi,LB-Kim,EamonnIdxDTW05,Lemire-FastDTW}.

Divide and conquer (DC) heuristic proposed by~\citet{Hirschberg75};
that is a dynamic programming algorithm that finds the least  cost
sequence alignment between two strings in linear space and quadratic
time. The algorithm was first used in speech recognition area to
solve the Longest Common Subsequence (LCSS). However as we will show
with the help of an example, \emph{DC} does not guarantee the
optimality of the \emph{DTW} distance.

~\citet{DTW-Originator} speed up the \emph{DTW} by constraining the
warping path to lie within a band around the diagonal. However, if
the optimal path crosses the band, the result will not be optimal.

~\citet{EamonnIdxDTW05} and~\citet{Lemire-FastDTW} introduced efficient lower bounds that
reduce the number of \emph{DTW} computations in a time series
database context. However, these lower bounds do not reduce the space
complexity of the \emph{DTW} computation, which is the objective of our
work.

    ~\citet{FTW05} presented FTW, a search method for DTW; it
    adds no global constraints on DTW. Their method designed
    based on a lower bounding distance measure that approximates the DTW
    distance. Therefore, it minimizes the number of
    DTW computations but does not increase the speed the DTW itself.

    ~\citet{FastDTW04} introduced an approximation algorithm for
    \emph{DTW} called \emph{FastDTW}. Their algorithm begins by using
    \textit{DTW} in  very low  resolution, and progresses to a
    higher resolution linearly in space and time. \emph{FastDTW} is
    performed in three steps: coarsening shrinks the time series
    into a smaller time series; the time series is projected by
    finding the minimum distance (warping path) in the lower
    resolution; and the warping path is an initial step for higher resolutions.
    The authors refined the warping path using local adjustment.
    \emph{FastDTW} is an approximation algorithm, and thus there is
    no guarantee it will always find the optimal path. It
    requires the coarsening step to be run several times to produce many
    different resolutions of the time series. The \emph{FastDTW}
    approach depends on a radius parameter as a constraint on the
    optimal path; however, our technique does not place any constrain
    while calculating the \emph{DTW} distance.

    \emph{DTW} has been used in data streaming
    problems.~\citet{SDTW-2007} proposed a new technique,
    Stream-DTW (\emph{STDW}). This measure is a lower bound of the
    \emph{DTW}. Their method uses a sliding window of size 512. They
    incorporated a band constraint, forcing the path to
    stay within the band frontiers, as in~\citep{DTW-Originator}.

 All the above algorithms were proposed
either to speed up \emph{DTW}, by reducing its space and time
complexity, or reducing the number of \emph{DTW} computations.
Interestingly, the approach of exploiting the similarity between
points (correlation) has never, to the best of our knowledge, been
used in finding the optimality between two time series.
\emph{SparseDTW} considers the correlation between data points, that
allows us to use a sparse matrix to store the warping matrix instead
of a full matrix. We do not believe that the idea of sparse matrix
has been considered previously to reduce the required space.


\begin{figure*}[t]
  \begin{center}
    \mbox{
            \subfigure[The alignment of measurements for measuring the DTW distance between
                        the two sequences $S$ and $Q$.\label{Fig:TSM-WarpingA}]
                       {\includegraphics [height=2.5in,width=0.5\textwidth]{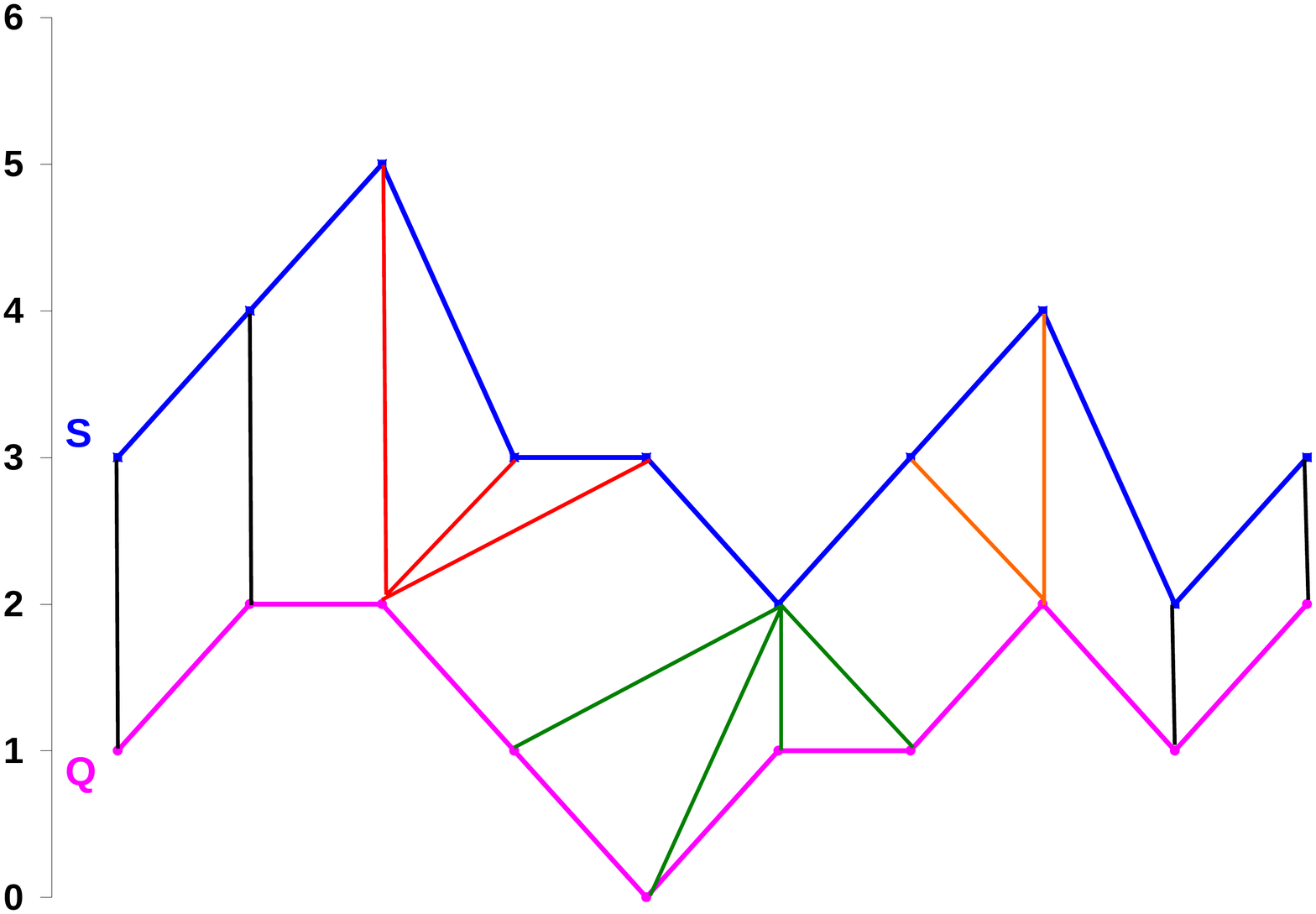}}
            \subfigure[The warping matrix $D$ produced by DTW; highlighted cells constitute the optimal warping path.\label{Fig:TSM-WarpingB}]
                      {\includegraphics [height=2.5in,width=0.5\textwidth]{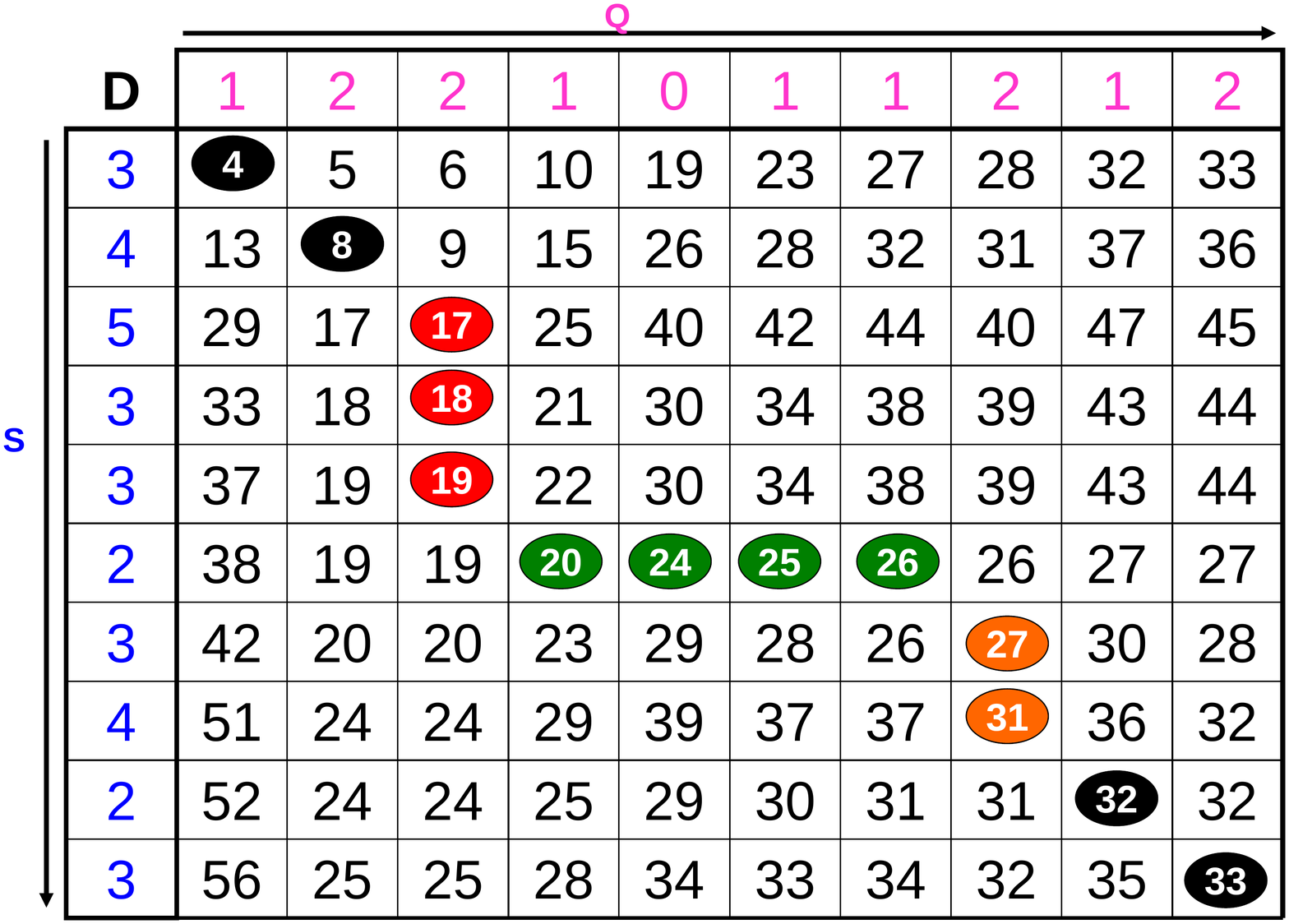}}
        }
      \caption{Illustration of \emph{DTW}.}
    \label{Fig:TSM-Warping}
  \end{center}
\end{figure*}


\begin{algorithm}[thb]
 \caption{\emph{DTW}: The standard DTW algorithm.} \label{Alg:TSM-DTW}
\begin{algorithmic}[1]

\REQUIRE \emph{$S$: Sequence of length $n$},
 \emph{$Q$: Sequence of length $m$.}

\ENSURE \emph{DTW distance.}

\STATE \emph{Initialize} $D(i,1) \Leftarrow i \delta$ \emph{for each
i}

\STATE \emph{Initialize} $D(1,j) \Leftarrow j \delta$ \emph{for each
j}

\FORALL{$i$ such that $2\leq i\leq n$}

    \FORALL{$j$ such that $2\leq j\leq m$}
        \STATE \emph{Use Equation~\ref{Equ:TSM-DTW} to compute $D(i,j)$}
    \ENDFOR

\ENDFOR

\RETURN $D(n,m)$

\end{algorithmic}
\end{algorithm}

\section{Dynamic Time Warping (DTW)} \label{Sec:TSM-DTWBackground}

DTW is a dynamic programming technique used for measuring the
similarity between any two time series with arbitrary lengths. This
section gives an overview of \emph{DTW} and how it is calculated.
The following two time series (Equations~\ref{Equ:TSM-Seq1}
and~\ref{Equ:TSM-Seq2}) will be used in our explanations.
 \begin{eqnarray} \label{Equ:TSM-Seqs}
   S &=& s_{1}, s_{2}, s_{3}, \cdots, s_{i}, \cdots, s_{n}\label{Equ:TSM-Seq1}\\
   Q &=& q_{1}, q_{2}, q_{3}, \cdots, q_{j}, \cdots, q_{m}\label{Equ:TSM-Seq2}
 \end{eqnarray}

 Where $n$ and $m$ represent the length of time series $S$ and $Q$, respectively. $i$ and $j$ are the point indices in the time series.

\emph{DTW} is a time series association algorithm that was
originally used in speech recognition \citep{DTW-Originator}. It
relates two time series of feature vectors by warping the time axis
of one series onto another.

As a dynamic programming technique, it divides the problem into
several sub-problems, each of which contribute in calculating the
distance cumulatively. Equation~\ref{Equ:TSM-DTW} shows the
recursion that governs the computations is:

\begin{equation}\label{Equ:TSM-DTW}
    D(i,j)= d(i,j)+min \left\{ \begin{array}{l}
                                D(i-1,j)\\
                                D(i-1,j-1)\\
                                D(i,j-1).
                            \end{array}
                    \right.
\end{equation}

The first stage in the \emph{DTW} algorithm is to fill a local
distance matrix $d$. That matrix has $n \times m$ elements which
represent the Euclidean distance between every two points in the
time series (i.e., distance matrix). In the second stage, it fills
the warping matrix $D$ (Figure~\ref{Fig:TSM-WarpingB}) on the basis
of Equation~\ref{Equ:TSM-DTW}. Lines 1 to 7 in
Algorithm~\ref{Alg:TSM-DTW} illustrate the process of filling the
warping matrix. We refer to the cost between the $i^{th}$ and the
$j^{th}$ elements as $\delta$ as mentioned in line 1 and 2.

After filling the warping matrix, the final stage for the \emph{DTW}
is to report the optimal warping path and the \emph{DTW} distance.
Warping path is a set of adjacent matrix elements that identify the
mapping between $S$ and $Q$. It represents the path that minimizes
the overall distance between $S$ and $Q$. The total number of
elements in the warping path is $K$, where $K$ denotes the
normalizing factor and it has the following attributes:

   \[W= w_{1}, w_{2}, \ldots, w_{K}\]
   \[ max(|S|,|Q|)\leq K < (|S|+|Q|)\]

Every warping path must satisfy the following constraints
\citep{EamonnIdxDTW05,FastDTW04,DTW-Originator}:

\begin{enumerate}
  \item \textbf{Monotonicity:} Any two adjacent elements of the warping path $W$,  $w_{k}=(w_{i},w_{j})$ and
    $w_{k-1}=(w_{i}^{'},w_{j}^{'})$, follow the inequalities, $w_{i}-w_{i}^{'}\geq 0$ and $w_{j}-w_{j}^{'}\geq 0$.
    This constrain guarantees that the warping path will not roll back on itself. That is, both indexes
    $i$ and $j$ either stay the same or increase (they never
    decrease).
  \item \textbf{Continuity:} Any two adjacent elements of the warping
    path $W$,$w_{k}=(w_{i},w_{j})$ and $w_{k+1}=(w_{i}^{'},w_{j}^{'})$, follow the inequalities, $w_{i}-w_{i}^{'}\leq
    1$ and $w_{j}-w_{j}^{'}\leq 1$. This constraint guarantees that the warping path advances one step at a time.
    That is, both indexes $i$ and $j$ can only increase by at most 1 on each step along the
    path.
  \item \textbf{Boundary:} The warping path starts from the top left
    corner $w_{1}=(1,1)$ and ends at the bottom right corner
    $w_{k}=(n,m)$. This constraint guarantees that the warping path
    contains all points of both time series.
\end{enumerate}

Although there are a large number of warping paths that satisfy all
of the above constraints, \emph{DTW} is designed to find the one
that minimizes the warping cost (distance).
Figures~\ref{Fig:TSM-WarpingA} and~\ref{Fig:TSM-WarpingB}
demonstrate an example of how two time series ($S$ and $Q$) are
warped and the way their distance is calculated. The circled cells
show the optimal warping path, which crosses the grid from the top
left corner to the bottom right corner. The \emph{DTW} distance
between the two time series is calculated based on this optimal
warping path using the following equation:

\begin{equation}\label{Equ:TSM-WarpingPath}
    DTW(S,Q)= min\left\{\frac{\sqrt{\sum_{k=1}^{K}W_{k}}}{K}
                 \right.
\end{equation}

 The $K$ in the denominator is used to normalize different warping paths with different lengths.

Since the \emph{DTW}  has to potentially examine every cell in the
warping matrix, its space and time complexity is $O(nm)$.

\begin{figure}
    \begin{center}
     \includegraphics [height=3in,width=0.95\columnwidth]{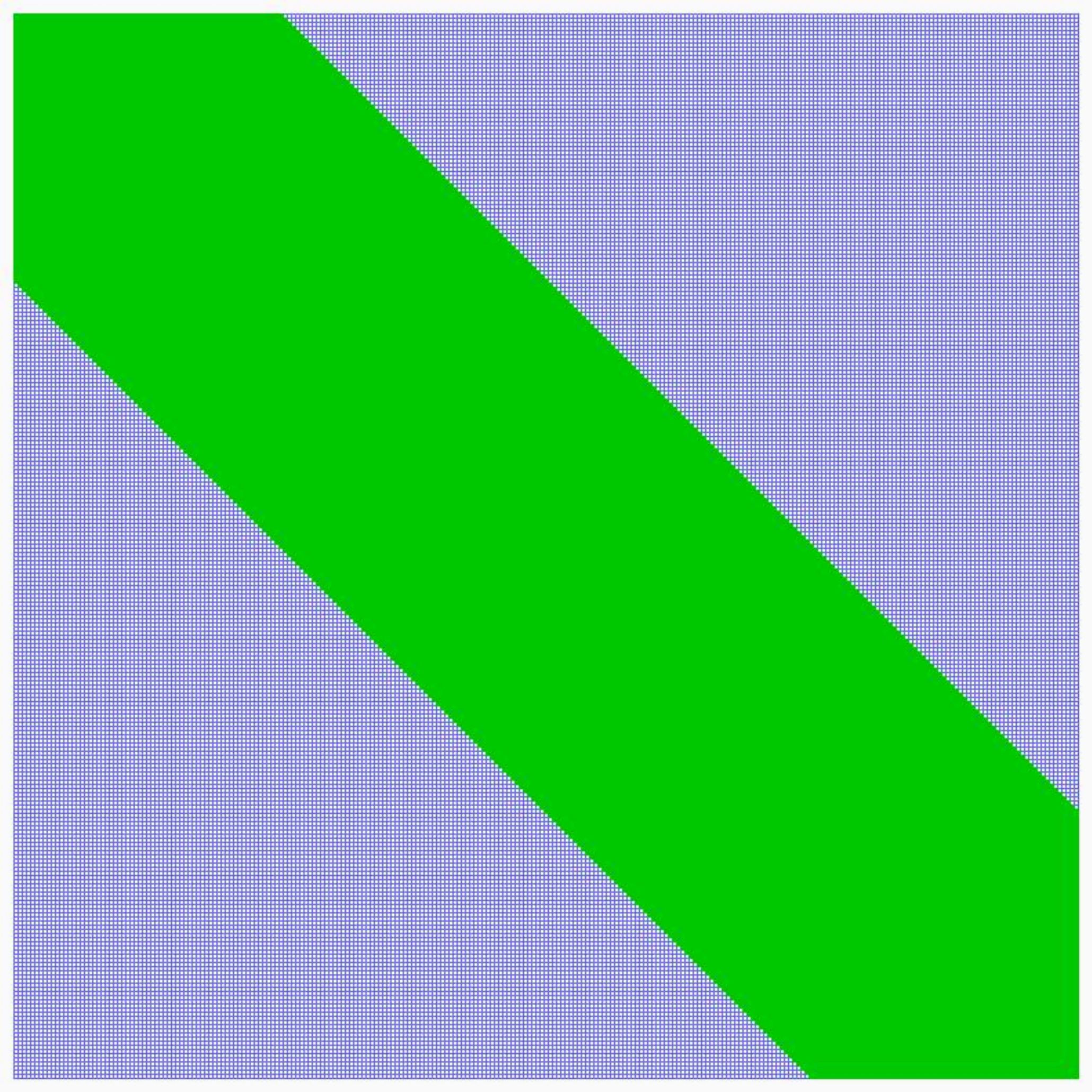}
     \caption{Global constraint (Sakoe Chiba Band), which limits the
              warping scope. The diagonal green areas correspond to the warping scopes.}
     \label{Fig:TSM-SakoeChibaBand}
    \end{center}
 \end{figure}

\section{Global Constraint (BandDTW)} \label{Sec:TSM-BandDTW}

There are several methods that add global constraints on DTW to
increase its speed by limiting how far the warping path may stray
from the diagonal of the warping
matrix~\citep{Tappert78,Berndt94,Cory80}. In this paper we use
Sakoe-Chiba Band (henceforth, we refer to it as
BandDTW)~\citet{DTW-Originator} when comparing with our proposed
algorithm (Figure~\ref{Fig:TSM-SakoeChibaBand}). BandDTW used to
speed up the \emph{DTW} by adding constraints which force the
warping path to lie within a band around the diagonal; if the
optimal path crosses the band, the DTW distance will not be optimal.

\begin{algorithm}[thb]

 \caption{\emph{DC}: Divide and Conquer technique.} \label{Alg:TSM-DC}
\begin{algorithmic}[1]

\REQUIRE \emph{$S$: Sequence of length $n$},
 \emph{$Q$: Sequence of length $m$.}

\ENSURE \emph{DTW distance.}

\STATE \emph{Divide-Conquer-Alignment(S,Q)}

\STATE $n \Leftarrow |S|$

\STATE $m \Leftarrow |Q|$

\STATE $Mid \Leftarrow \lceil m/2 \rceil$

\IF  {$n \leq 2$ or $m \leq 2$}

\STATE \emph{Compute optimal alignment using standard DTW}

\ELSE

\STATE $f \Leftarrow$ \emph{ForwardsSpaceEfficientAlign(S,Q[1:Mid])}

\STATE $g \Leftarrow$\emph{BackwardsSpaceEfficientAlign(S,Q[Mid:m])}

\STATE $q \Leftarrow$ \emph{index that minimizing}
$f(q,Mid)+g(q,Mid)$

\STATE \emph{Add (q,Mid) to global array P}

\STATE \emph{Divide-Conquer-Alignment(S[1:q],Q[1:Mid])}

\STATE \emph{Divide-Conquer-Alignment(S[q:n],Q[Mid:m])}

\ENDIF

\RETURN \emph{P}

\end{algorithmic}
\end{algorithm}

 \begin{figure*}[thb]
    \begin{center}
     \includegraphics[width=\textwidth]{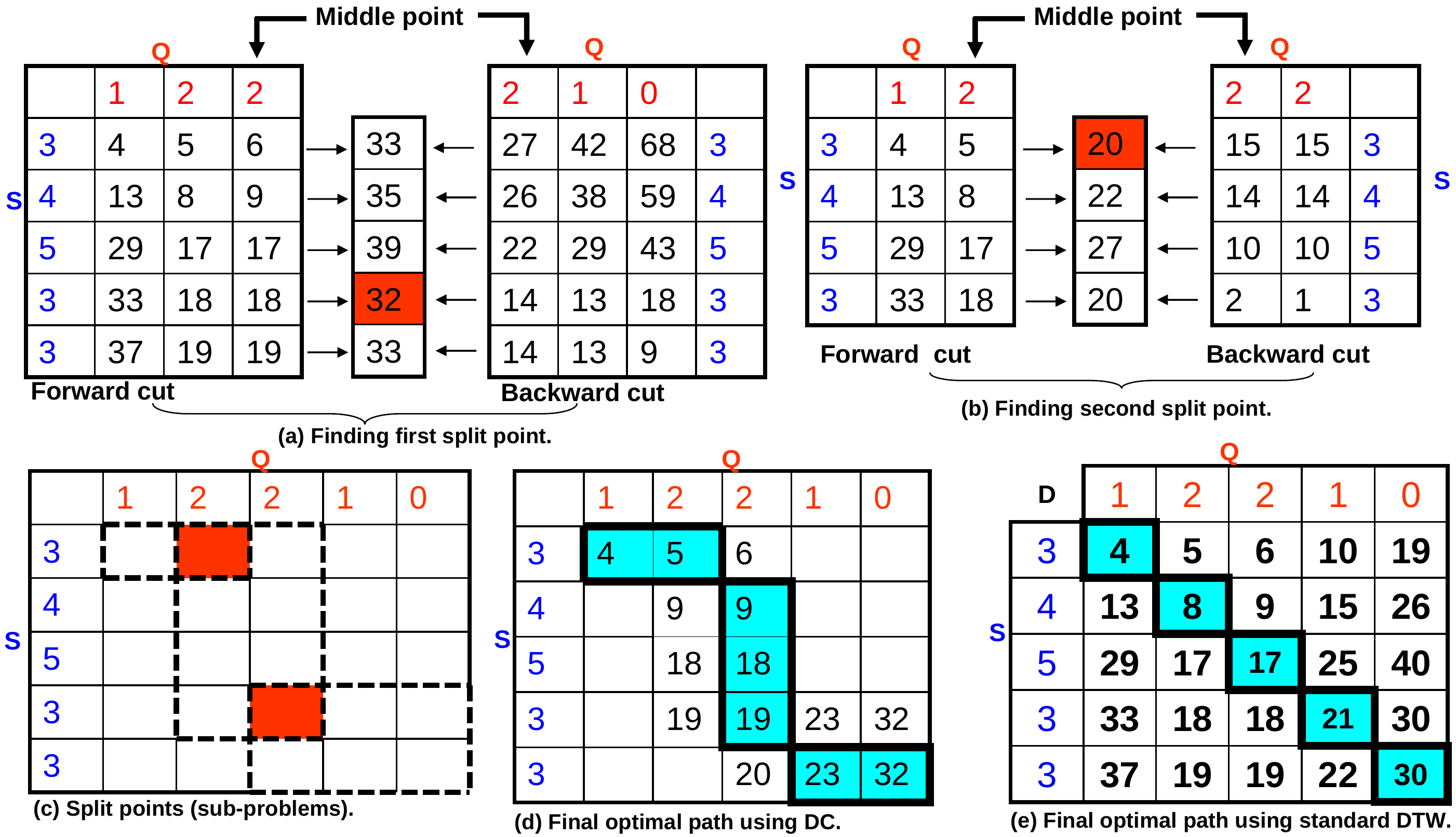}
     \caption{An example to show the difference between the standard
     DTW and the DC algorithm.}
     \label{Fig:TSM-DTW-DC-Example}
    \end{center}
 \end{figure*}

\section{Divide and Conquer Technique (DC)} \label{Sec:TSM-DC}

In the previous section, we have shown how to compute the optimal
alignment using the standard \emph{DTW} technique between two time
series. In this section we will show another technique that uses a
Divide and Conquer heuristic, henceforth we refer to it as
(\emph{DC}), proposed by~\citet{Hirschberg75}. \emph{DC} is a
dynamic programming algorithm used to find the least cost sequence
alignment between two strings. The algorithm was first introduced to
solve the Longest Common Subsequence (LCSS) \citep{Hirschberg75}.
Algorithm~\ref{Alg:TSM-DC} gives a high level description of the
\emph{DC} algorithm. Like in the standard sequence alignment, the
\emph{DC} algorithm has $O(mn)$ time complexity but $O(m+n )$ space
complexity, where $m$ and $n$ are the lengths of the two sequences
being aligned. We will be using Algorithm~\ref{Alg:TSM-DC}  along
with Figure~\ref{Fig:TSM-DTW-DC-Example} to explain how \emph{DC}
works. In the example we use two sequences $S=[3,4,5,3,3]$ and
$Q=[1,2,2,1,0]$ to determine the optimal alignment between them.
There is only one optimal alignment for this example
(Figure~\ref{Fig:TSM-DTW-DC-Example}(e)), where shaded cells are the
optimal warping path. The \emph{DC} algorithm works as follows:

\begin{enumerate}
  \item It finds the middle point in $Q$ which is $Mid=|Q|/2$,
    (Figure~\ref{Fig:TSM-DTW-DC-Example}(a)). This helps to find the
    split point which divides the warping matrix into two parts
    (sub-problems). A forward space efficiency function (Line 8) uses
    $S$ and the first cut of $Q=[1,2,2]$, then a backward step (Line 9)
    uses $S$ and $Q=[2,1,0]$ (Figure~\ref{Fig:TSM-DTW-DC-Example}(a)). Then by adding the last column from the
    forward and backward steps together and finding the index of the
    minimum value, the resultant column indicates the row index that
    will be used along with the middle point to locate the split point
    (shaded cell in Figure~\ref{Fig:TSM-DTW-DC-Example}(a)). Thus, the first
    split point is D(4,3). At this stage of the algorithm, there are two
    sub-problems; the alignment of $S=[3,4,5,3]$ with $Q=[1,2,2]$ and of
    $S=[3,3]$ with $Q=[2,1,0]$.

  \item \emph{DC} is recursive algorithm, each call splits the
    problem into two other sub-problems if both sequences are of
    $length>2$, otherwise it calls the standard \emph{DTW} to find the
    optimal path for that particular sub-problem. In the example, the
    first sub-problem will be fed to Line 12 which will find another
    split point, because both input sequences are of length $>2$. Figure~\ref{Fig:TSM-DTW-DC-Example}(b) shows how the new split point is found.
    Figure~\ref{Fig:TSM-DTW-DC-Example}(c) shows the two split points (shaded
    cells) which yield to have sub-problems of sequences of length
    $\leq2$. In this case \emph{DTW} will be used to find the optimal
    alignment for each sub-problem.

  \item The \emph{DC} algorithm finds the final alignment by
    concatenating the results from each call of the standard
    \emph{DTW}.
\end{enumerate}

The example in Figure~\ref{Fig:TSM-DTW-DC-Example} clarifies that
the \emph{DC} algorithm does not give the optimal warping path.
Figures~\ref{Fig:TSM-DTW-DC-Example}(d) and (e) show the paths
obtained by the \emph{DC} and \emph{DTW} algorithms, respectively.

\emph{DC} does not yield the optimal path as it goes into infinite
recursion because of how it calculates the middle point. \emph{DC}
calculates the middle point as follows:

There are two scenarios: first, when the middle point
(Algorithm~\ref{Alg:TSM-DC} Line 4) is floored ($Mid= \lfloor m/2
\rfloor$) and second when it is rounded up ($Mid= \lceil m/2
\rceil$). The first scenario causes infinite recursion, since the
split from the previous step gives the same sub-sequences (i.e., the
algorithm keeps finding the same split point). The second scenario
is shown in Figures~\ref{Fig:TSM-DTW-DC-Example}(a-d), which clearly
confirms that the final optimal path is not the same as the one
retrieved by the standard \emph{DTW} \footnote{It should be noted
that our example has only one optimal path that gives the optimal
distance.}. The final \emph{DTW} distance is different as well. The
shaded cells in Figures~\ref{Fig:TSM-DTW-DC-Example}(d) and (e) show
that both warping paths are different.

\section{Sparse Dynamic Programming Approach} \label{Sec:TSM-SparseDTW}

In this section, we outline the main principles we use in
\emph{SparseDTW} and follow up with an illustrated example along
with the \emph{SparseDTW} pseudo-code. We exploit the following
facts in order to reduce space usage while avoiding any
re-computations:

\begin{enumerate}
  \item Quantizing the input time series to exploit the similarity between the points in the two time series.
  \item Using a sparse matrix of size $k$, where $k=n \times m$ in the worst case. However, if the two sequences are similar,
        $k$ $<< n \times m$.
  \item The warping matrix is calculated using dynamic programming and sparse matrix indexing.
\end{enumerate}

\subsection{Key Concepts}

In this section we introduce the key concepts used in our algorithm.

\begin{figure*}[t]
    \begin{center}
     \includegraphics[width=\textwidth]{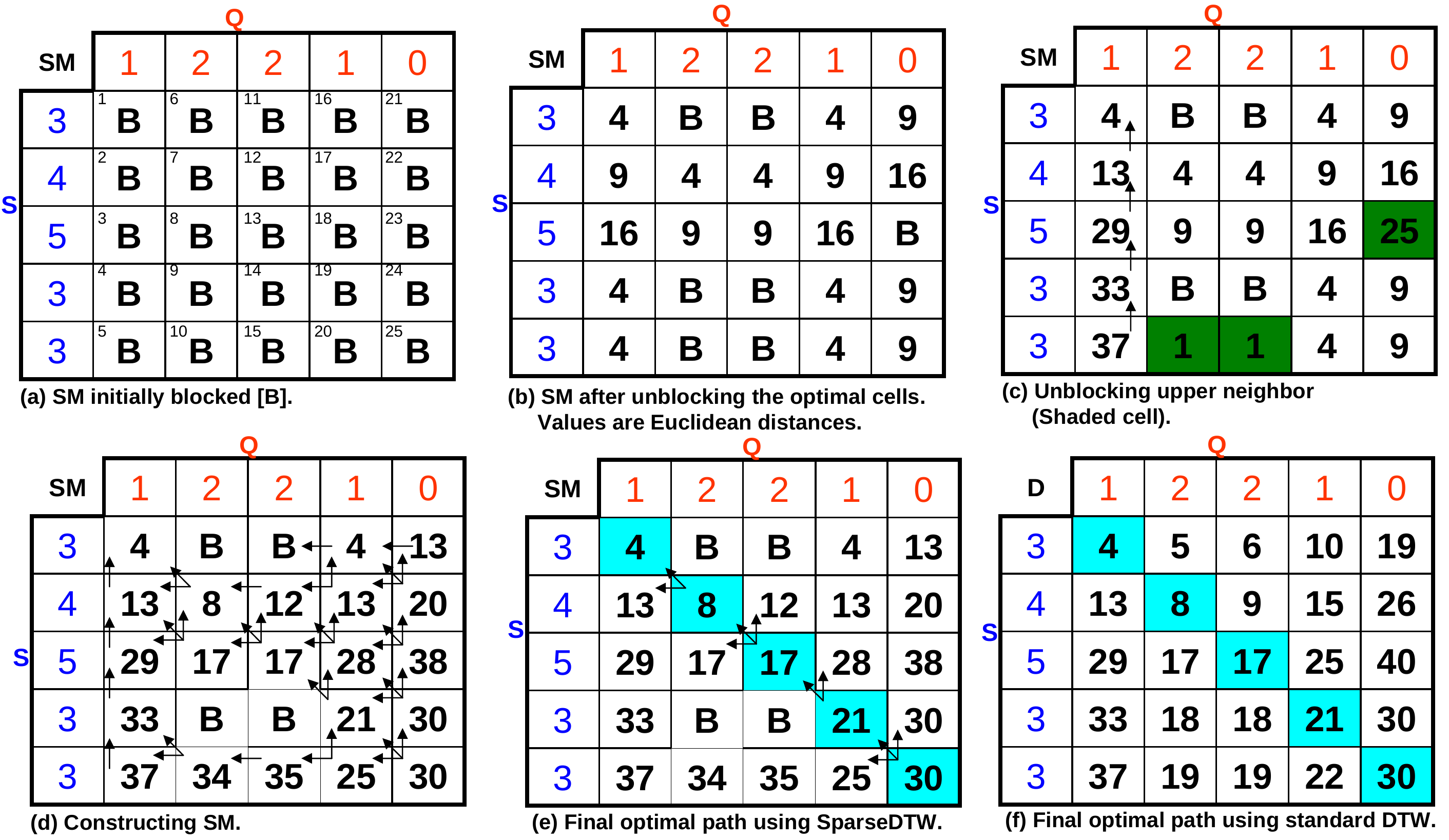}
     \caption{An example of the SparseDTW algorithm and the method of finding the optimal path.}
     \label{Fig:TSM-Example-New}
    \end{center}
 \end{figure*}

\begin{definition} [Sparse Matrix $SM$]
     is a matrix that is populated largely with zeros. It allows the techniques to take advantage of
     the large number of zero  elements. Figure~\ref{Fig:TSM-Example-New}(a) shows the $SM$ initial state.
     $SM$ is linearly indexed, The little numbers, in the top left corner of $SM$'s cells, represent the
     cell index. For example, the indices of the cells $SM(1,1)$ and $SM(5,5)$ are 1 and 25, respectively.
\end{definition}

\begin{definition} [Lower Neighbors ($LowerNeighbors$)]

a cell $c \in SM$ has three lower neighbors which are the cells of
the indices ($c-1$), ($c-n$), and ($c-(n+1)$) (where $n$ is the
number of rows in $SM$) . For example, the lower neighbors of cell $SM(12)$ are $SM(6)$, $SM(7)$ and $SM(11)$
(Figure~\ref{Fig:TSM-Example-New}(a)).

\end{definition}


\begin{definition} [Upper Neighbors ($UpperNeighbors$)]
 a cell $c \in SM$ has three upper neighbors which are the cells of the indices ($c+1$), ($c+n$),
and ($c+n+1$) (where $n$ is the number of rows in $SM$) . For
example, the upper neighbors of cell $SM(12)$ are $SM(13)$, $SM(17)$ and $SM(18)$ (Figure~\ref{Fig:TSM-Example-New}(a)).

\end{definition}

\begin{definition}[Blocked Cell (B)]
  a cell $c \in SM$  is blocked if its value is zero. The letter (B)
  refers to the blocked cells (Figure~\ref{Fig:TSM-Example-New}(a)).
\end{definition}

\begin{definition}[Unblocking]

Given a cell $c \in SM$, if $SM(c)$'s upper  neighbors
($SM(c+1)$,$SM(c+n)$, and $SM(c+n+1)$) are blocked, they will be
unblocked. Unblocking is performed  by calculating the EucDist for
these cells and adding them to $SM$. In other words, adding the
distances to these cells means changing  their state  from blocked
(B) into unblocked. For example, $SM(10)$ is a blocked upper neighbor of $SM(5)$, in this case $SM(10)$ needs to be unblocked (Figure~\ref{Fig:TSM-Example-New}(c)).

\end{definition}

\subsection{SparseDTW Algorithm} \label{Sec:TSM-SparseDTWAlgo}

Algorithm~\ref{Alg:TSM-SparseDTW} takes $Res$, the resolution parameter as an input that
determines the number of bins as $\frac{2}{Res}$. $Res$ will have no
impact on the optimality. We now present an example of our algorithm
to illustrate some of the highlights of our approach: We start with
two sequences:

$S=[3,4,5,3,3]$ and $Q=[1,2,2,1,0]$.

In Line 1, we first quantize the sequences into the range $[0,1]$
using Equation~\ref{Equ:TSM-Quantization}:

       \begin{equation}\label{Equ:TSM-Quantization}
             QuantizedSeq_{i}^{k}=\frac{S_{i}^{k}-min(S^{k})}{max(S^{k})-min(S^{k})}.
        \end{equation}

        Where $S_{i}^{k}$ denotes the $i^{th}$ element of the
        $k^{th}$ time series. This yields the following sequences:

$S'=[0,0.5,1.0,0.0,0.0]$ and  $Q'=[0.5,1.0,1.0,0.5,0]$

In Lines 4 to 7 we create {\it overlapping} bins, governed by two
parameters: bin-width and the overlapping width (which we refer to
as the resolution). It is important to note that these two
parameters do not affect the optimality of the alignment but do have
an affect on the amount of space utilized. For this particular
example, the bin-width is 0.5. We thus have 4 bins which are
shown in Table~\ref{Tab:TSM-Bins}.\\

\begin{table}[thb]
 \centering
    \begin{tabular}{clll}
     \hline \hline
      Bin Number   &  Bin         & Indices & Indices \\
      ($B_{k}$)    &  Bounds       & of $S'$ &of $Q'$  \\
      \hline\hline
         1    & 0.0-0.5   & 1,2,4,5  & 1,4,5    \\\hline
         2    & 0.25-0.75 & 2        & 1,4      \\\hline
         3    & 0.5-1.0   & 2,3      & 1,2,3,4  \\\hline
         4    & 0.75-1.25  & 3        & 2,3      \\\hline\hline
    \end{tabular}
 \caption{Bins bounds, where $B_k$ is the $k^{th}$ bin.}\label{Tab:TSM-Bins}
\end{table}

Our intuition is that points in sequences with similar profiles will
be mapped to other points in the same bin or neighboring bins. In
which case the non-default entries of the sparse matrix can be used
to compute the warping path. Otherwise, default entries of the
matrix will have to be ``opened'', reducing the sparsity of the
matrix but never sacrificing the optimal alignment.

In Lines 3 to 13, the sparse warping matrix $SM$ is constructed
using the equation below. $SM$\footnote{If the Euclidean distance
(EucDist) between $S(i)$ and $Q(j)$ is zero, then $SM(i,j) = -1$, to
distinguish between a blocked cell and any cell that represents zero
distance.} is a matrix that has generally few non-zero (or
``interesting") entries. It can be represented in much less than $n
\times m$ space, where $n$ and $m$ are the lengths of the time
series $S$ and $Q$, respectively.

\begin{equation}\label{Equ:TSM-SM}
 \centering
    SM(i,j) = \left
                \{\begin{array}{ll}
                        EucDist(S(i),Q(j)) &  \mbox{\small{if $S(i)$ and $Q(j) \in B_{k}$}}  \\
                        B                & otherwise
                 \end{array}
            \right.
\end{equation}

We assume that $SM$ is linearly ordered and the default value of
$SM$ cells are zeros. That means the cells initially are
\emph{Blocked} (B) (Figure~\ref{Fig:TSM-Example-New}(a)).
Figure~\ref{Fig:TSM-Example-New}(a) shows the linear order of the
$SM$ matrix, where the little numbers on the top left corner of each
cell represent the index of the cells. In Line 6 and 7, we find the
index of each quantized value that falls in the bin bounds
(Table~\ref{Tab:TSM-Bins} column 2, 3 and 4). The
Inequality~\ref{Euq:LowerUpperBounds} is used in Line 6 and 7 to
find the indices of the default entries of the $SM$.

\begin{equation}\label{Euq:LowerUpperBounds}
    LowerBound \leq QuantizedSeq^{k}_{i} \leq UpperBound.
\end{equation}

Where $LowerBound$ and $UpperBound$ are the bin bounds and
$QuantizedSeq^{k}_{i}$ represents the quantized time series which
can be calculated using Equation~\ref{Equ:TSM-Quantization}.

Lines 8 to 12 are used to initialize the $SM$. That is by joining
all indices in $idxS$ and $idxQ$ to open corresponding cells in
$SM$. After unblocking (opening) the cells that reflect the
similarity between points in both sequences, the $SM$ entries are
shown in Figure~\ref{Fig:TSM-Example-New}(b).

Lines 14 to 22 are used to calculate the warping cost. In Line 15,
we find the warping cost for each open cell $c \in SM$ (cell $c$ is
the number from the linear order of $SM$'s cells) by finding the
minimum of the costs of its lower neighbors, which are
$[c-1,c-n,c-(n+1)]$ (black arrows in
Figure~\ref{Fig:TSM-Example-New}(d) show the lower neighbors of
every open cell). This cost is then added to the local distance of
cell $c$ (Line 17). The above step is similar to \textit{DTW},
however, we may have to open new cells if the upper neighbors at a
given local cell $c \in SM$ are blocked. The indices of the upper
neighbors are $[c+1,c+n,c+n+1]$ , where $n$ is the length of
sequence $S$ (i.e., number of rows in $SM$). Lines 18 to 21 are used
to check always the upper neighbors of $c \in SM$. This is performed
as follows:  if the $|UpperNeighbors|=0$ for a particular cell, its
upper neighbors will be unblocked. This is very useful when the
algorithm traverses $SM$ in reverse to find the final optimal path.
In other words, unblocking allows the path to be connected. For
example, the cell $SM(5)$  has one upper neighbor that is cell
$SM(10)$ which is blocked (Figure~\ref{Fig:TSM-Example-New}(b)),
therefore this cell will be unblocked by calculating the
EucDist(S(5),Q(2)). The value will be add to the $SM$ which means
that cell $SM(10)$ is now an entry in $SM$
(Figure~\ref{Fig:TSM-Example-New}(c)). Although unblocking adds
cells to $SM$ which means the number of open cells will increase,
but the overlapping in the bins boundaries allows the $SM$'s
unblocked cells to be connected mostly that means less number of
unblocking operations. Figure~\ref{Fig:TSM-Example-New}(d) shows the
final entries of the $SM$ after calculating the warping cost of all
open cells.

Lines 23 to 32 return the warping path. $hop$ initially represents
the linear index for the $(m,n)$ entry of $SM$, that is the bottom
right corner of $SM$ in Figure~\ref{Fig:TSM-Example-New}(e).
Starting from $hop=n \times m$ we choose the neighbors [$hop-n,
hop-1,hop-(n+1)$] with minimum warping cost and proceed recursively
until we reach the first entry of $SM$, namely $SM(1)$ or $hop=1$.
It is interesting that while calculating the warping path we only
have to look at the open cells, which may be fewer in number than 3.
This potentially reduces the overall time complexity.

Figure~\ref{Fig:TSM-Example-New}(e) demonstrates an example of how
the two time series ($S$ and $Q$) are warped and the way their
distance is calculated using \emph{SparseDTW}. The filled cells show
the optimal warping path, which crosses the grid from the top left
corner to the bottom right corner. The distance between the two time
series is calculated using Equation~\ref{Equ:TSM-WarpingPath}.
Figure~\ref{Fig:TSM-Example-New}(f) shows the standard \emph{DTW}
where the filled cells are the optimal warping path. It is clear
that both techniques give the optimal warping path which will yield
the optimal distance.


\begin{algorithm}[th]
 \caption{\emph{SparseDTW}: Sparse dynamic programming technique.}
\label{Alg:TSM-SparseDTW}
\begin{algorithmic}[1]

\REQUIRE \emph{$S$: Time series of length $n$},
 \emph{$Q$: Time series of length $m$,} and
 \emph{Res}.

\ENSURE \emph{Optimal warping path} and \emph{SparseDTW} distance.

\STATE $[S',Q'] \Leftarrow Quantize(S,Q)$

%

\STATE $LowerBound \Leftarrow 0$, $UpperBound \Leftarrow Res$

 \FORALL{$ 0 \leq LowerBound \leq 1-\frac{Res}{2}$}
        \STATE $IdxS \Leftarrow find (LowerBound \leq S' \leq UpperBound)$
        \STATE $IdxQ \Leftarrow find (LowerBound \leq Q' \leq UpperBound)$
        \STATE $LowerBound \Leftarrow LowerBound + \frac{Res}{2}$
        \STATE $UpperBound \Leftarrow LowerBound + Res$
        \FORALL{$idx_{i}\in IdxS$}
            \FORALL{$idx_{j}\in IdxQ$}
            \STATE Add $EucDist(idx_{i},idx_{j})$ to $SM$ \COMMENT{When $EucDist(idx_{i},idx_{j})=0, SM(i,j)=-1$.}
            \ENDFOR
        \ENDFOR
\ENDFOR

\COMMENT{Note: $SM$ is linearly indexed.}

 \FORALL{$c \in SM$}

    \STATE $LowerNeighbors \Leftarrow \{(c-1),(c-n),(c-(n+1))\}$
    \STATE $minCost \Leftarrow min(SM(LowerNeighbors))$ \COMMENT{SM(LowerNeighbors)=-1 means cost=0.}
    \STATE $SM(c) \Leftarrow SM(c) + minCost$

    \STATE $UpperNeighbors \Leftarrow \{(c+1),(c+n),(c+n+1)\}$
    \IF{$|UpperNeighbors|==0$}
        \STATE $SM \cup EucDist(UpperNeighbors)$
    \ENDIF
\ENDFOR

\STATE $WarpingPath \Leftarrow \Phi$

\STATE $hop \Leftarrow SM(n \times m)$ \COMMENT{Last index in SM.}

\STATE $WarpingPath \cup hop$

\WHILE{$hop \neq SM(1)$}
    \STATE $LowerNeighbors \Leftarrow \{(hop-1),(hop-n),(hop-(n+1))\}$
    \STATE [\emph{minCost,index}]$\Leftarrow$ \emph{min}[\emph{Cost}([\emph{LowerNeighbors}])
    \STATE $hop \Leftarrow index$
    \STATE $WarpingPath \cup hop$
\ENDWHILE

\STATE $WarpingPath \cup SM(1)$

\RETURN $WarpingPath, SM(n \times m)$

\end{algorithmic}
\end{algorithm}

\subsection{SparseDTW Complexity}

Given two time series $S$ and $Q$ of length $n$ and $m$,  the space
and time complexity of standard \textit{DTW} is $O(nm)$. For
\textit{SparseDTW} we attain a reduction by a constant factor $b$,
where $b$ is the number of bins. This is similar to the
\textit{BandDTW} approach where the reduction in space complexity is
governed by the size of the band. However, \textit{SparseDTW} always
yields the optimal alignment. The time complexity of
\textit{SparseDTW} is $O(nm)$ in the worst case as we potentially
have to access every cell in the matrix.

\section{Experiments, Results and Analysis} \label{Sec:TSM-Exps} In this
section we report and analyze the experiments that we have conducted
to compare \emph{SparseDTW} with other methods. Our main objective
is to evaluate the space-time tradeoff between \emph{SparseDTW},
\emph{BandDTW} and \emph{DTW}. We evaluate the effect of {\it
correlation} on the running time of \emph{SparseDTW}\footnote{The
run time includes the time used for constructing the Sparse Matrix
$SM$}. As we have noted before, both \emph{SparseDTW} and \emph{DTW}
always yield the optimal alignment while \emph{BandDTW} results can
often lead to sub-optimal alignments, as the optimal warping path
may lie outside the band. As we noted before \emph{DC} may not yield
the optimal result.

\subsection{Experimental Setup}

All experiments were carried out on a Windows XP operated PC with a
Pentium(R) D (3.4 GHz) processor and 2 GB main memory. The data
structures and algorithm were implemented in C++.

\subsection{Datasets}

We have used a combination of benchmark and synthetically generated
datasets. The benchmark dataset is a subset from the {\it UCR} time
series data mining archive \citep{UCR}. We have also generated
synthetic time series data to control and test the effect of
correlation on the running time of \emph{SparseDTW}. We briefly
describe the characteristics of each dataset used.

\begin{itemize}
  \item \textbf{GunX:} comes from the video surveillance
    application and captures the shape of a gun draw with the gun in
    hand or just using the finger. The shape is captured using 150 time
    steps and there are a total of 100 sequences \citep{UCR}. We randomly
    selected two sequences and computed their similarity using the three
    methods.

  \item \textbf{Trace:} is a synthetic dataset generated to
    simulate instrumentation failures in a nuclear power plant
    \citep{Roverso00multivariate}. The dataset consists of 200 time
    series each of length 273.

  \item \textbf{Burst-Water:} is formed by combining two different
    datasets from two different applications. The average length of the
    series is 2200 points \citep{UCR}.

  \item \textbf{Sun-Spot:} is a large dataset that has been
    collected since 1818. We have used the daily sunspot numbers. More
    details about this dataset exists in \citep{Sun-Spot}. The 1st column
    of the data is the year, month and day, the 2nd column is year and
    fraction of year (in Julian year)\footnote{The Julian year is a time
    interval of exactly 365.25 days, used in astronomy.}, and the 3rd
    column is the sunspot number. The length of the time series is 2898.

  \item \textbf{ERP:} is the Event Related Potentials that are
    calculated on human subjects\footnote{An indirect way of calculating
    the brain response time to certain stimuli}. The dataset consists of
    twenty sequences of length 256 \citep{Scott-ERP-Data}.

  \item \textbf{Synthetic:} Synthetic datasets were generated to
    control the correlation between sequences. The length of each
    sequence is 500.
\end{itemize}

 \begin{figure}
    \begin{center}
     \includegraphics[width=5in]{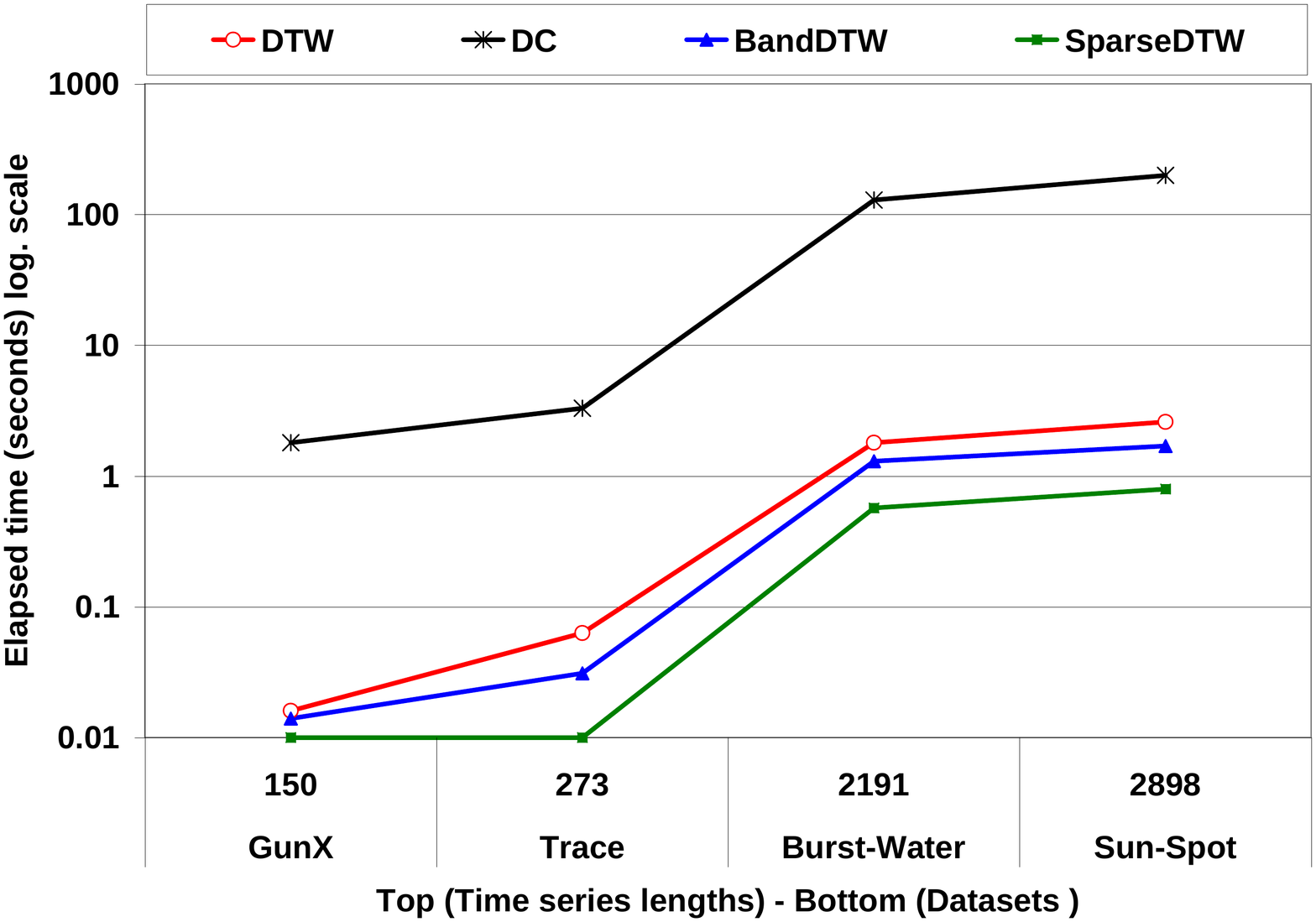}
     \caption{Elapsed time using real life datasets.}\label{Fig:TSM-ElapsedTime}
    \end{center}
 \end{figure}

\begin{figure}
    \begin{center}
     \includegraphics[width=5in]{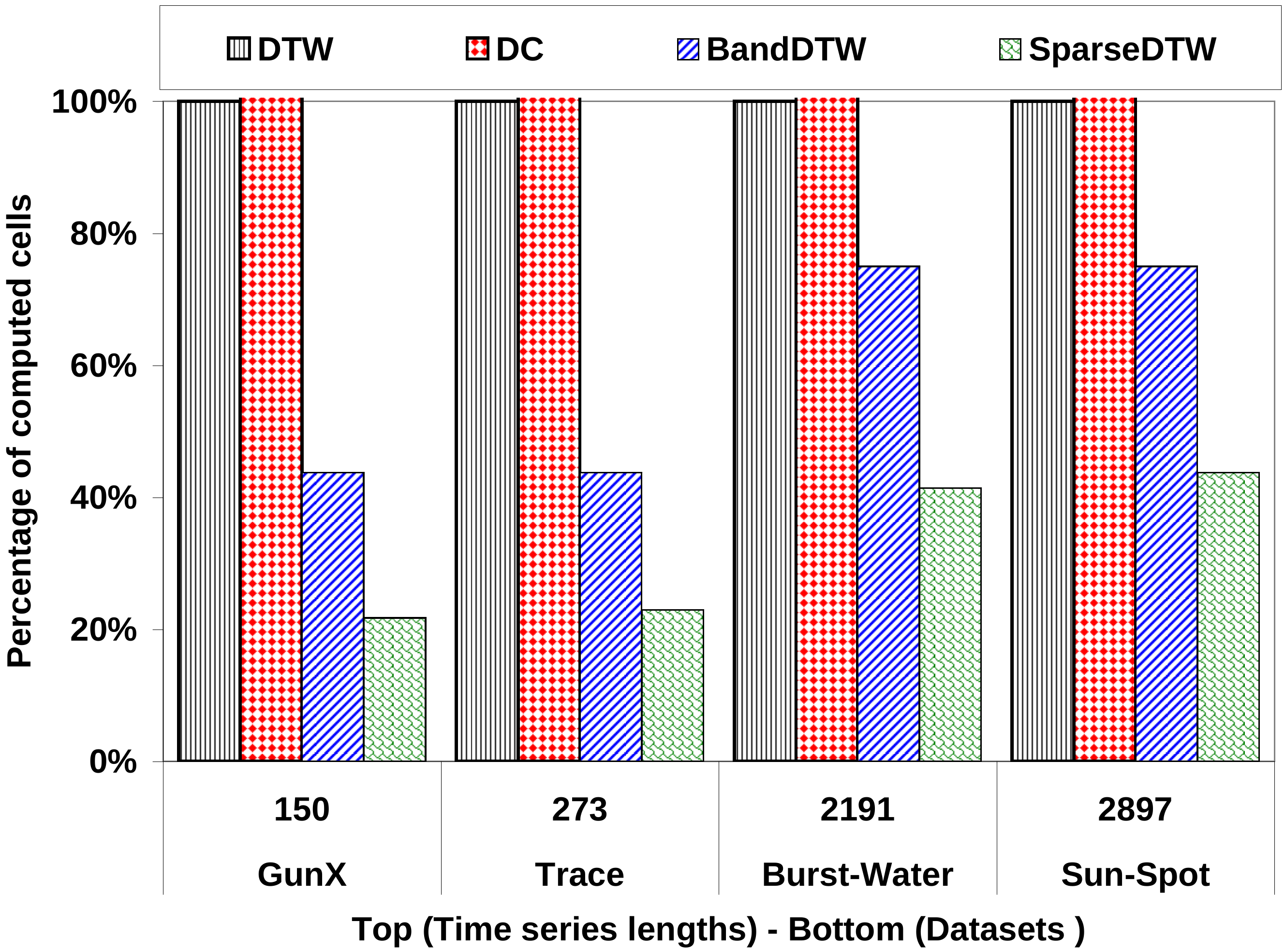}
     \caption{Percentage of computed cells as a measure for time complexity.}\label{Fig:TSM-ComputedCells}
    \end{center}
 \end{figure}

\begin{table} [thb]
\footnotesize \centering
\begin{tabular}{|l|r|r|r|r|}
    \hline
    & \multicolumn{4}{|c|}{\textbf{Number of computed cells used by} } \\
    \hline 
    \textbf{Data}&  &    	 &			&	\\
    \textbf{size}& \textbf{DTW}               & \textbf{DC}                &\textbf{BandDTW}&\textbf{SparseDTW} \\
    \hline\hline
    \textbf{2K}     & $4 \times 10^{6}$ & $>8 \times 10^{6}$& 2500   &2000 \\
    \hline
    \textbf{4K}     & $16 \times 10^{6}$ & $> 30 \times 10^{6}$&5000  &4000 \\
    \hline
    \textbf{6K}     & $36 \times 10^{6}$ & $> 70 \times 10^{6}$&7500  &6000 \\
    \hline
\end{tabular}
    \caption {\label{Tab:TSM-ComputedCells-Diagonal} Number of computed cells if the optimal path is close to the \emph{diagonal}.}
\end{table}


\begin{figure}
  \begin{center}
            \includegraphics [width=5in]{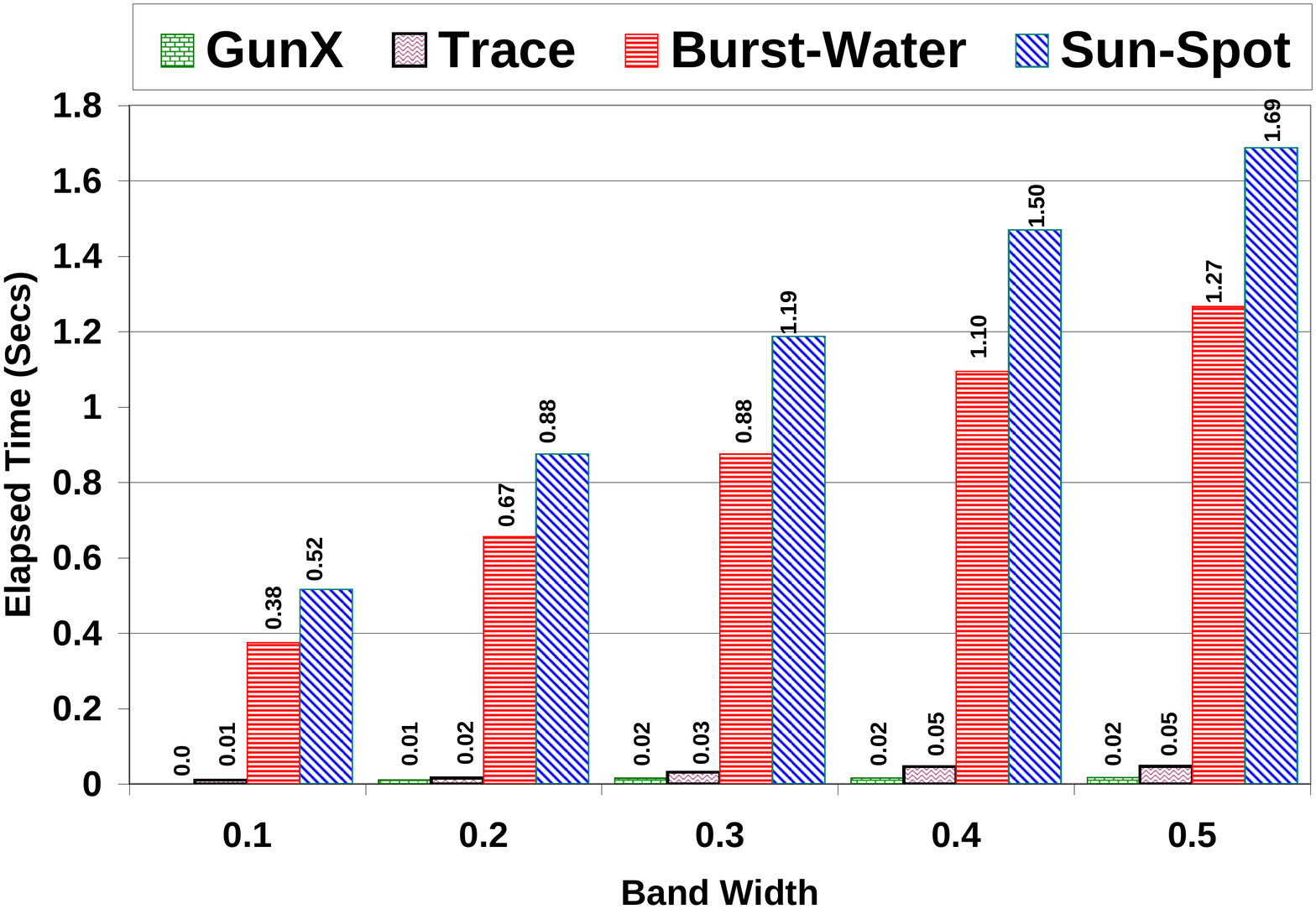}
      \caption{Effect of the band width on \emph{BandDTW} elapsed time.}
    \label{Fig:TSM-BandWidth}
  \end{center}
\end{figure}

\begin{figure}
  \begin{center}
       \includegraphics[width=5in]{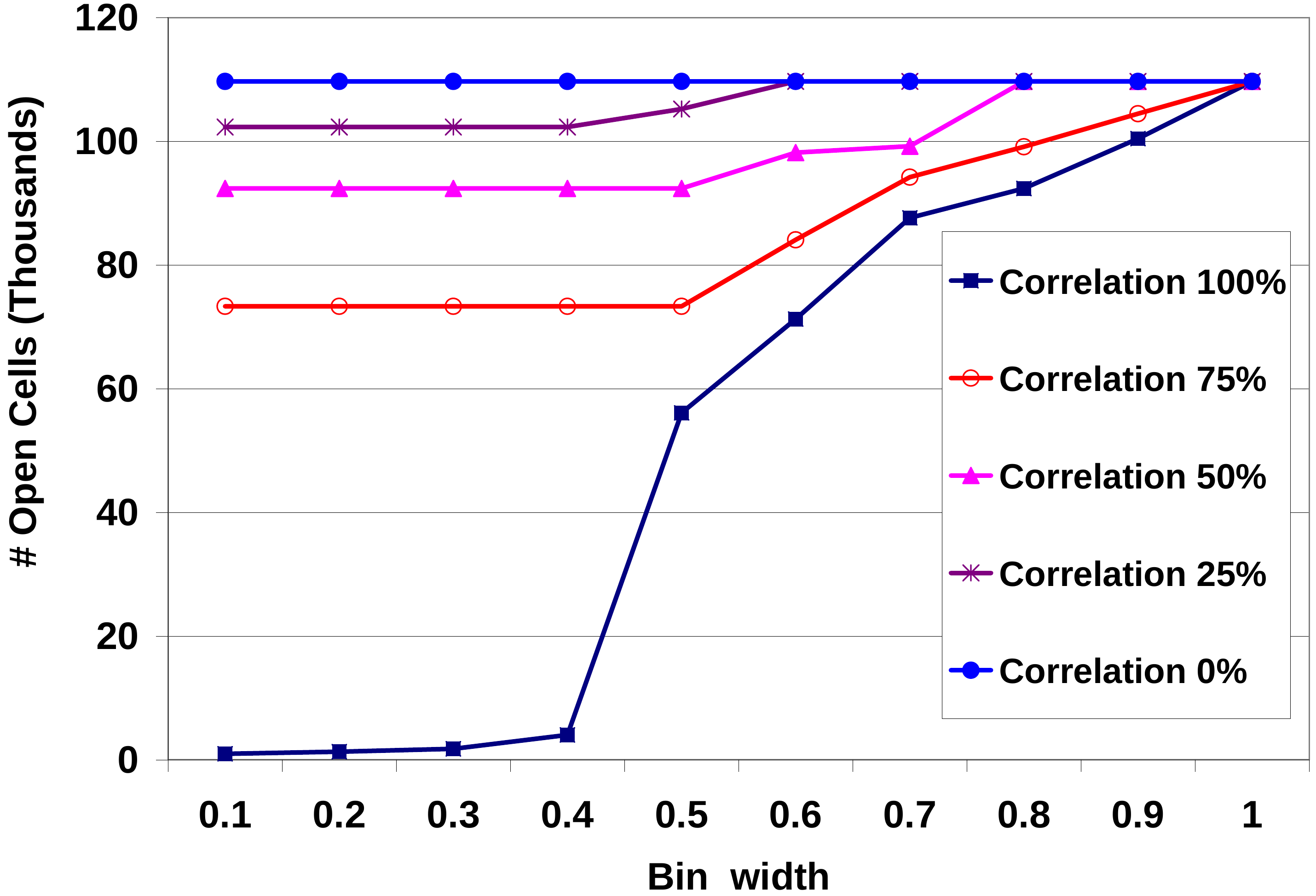} 
      \caption{Effects of the resolution and correlation on  \emph{SparseDTW}.}
    \label{Fig:TSM-Correlation}
  \end{center}
\end{figure}

\begin{table}
 \centering
    \begin{tabular}{l l r r } \\
        \hline \hline
        \textbf{Dataset}& \textbf{Algorithm}& \textbf{\#opened}&\textbf{Elapsed}\\
        \textbf{size}& \textbf{name}& \textbf{cells }&\textbf{Time(Sec.)}\\
        \hline\hline
        \multirow{2}{*}{\textbf{3K}} & DTW           & $9 \times 10^{6}$            & 7.3\\
                              & SparseDTW   & 614654            & 0.65\\
                              \hline
        \multirow{2}{*}{\textbf{6K}} & DTW           & $36 \times 10^{6}$            & 26\\
                              & SparseDTW   & 2048323           & 2.2\\
                              \hline
       \multirow{2}{*}{\textbf{9K}}&DTW              & $81 \times 10^{6}$            & N.A\\
                               & SparseDTW  & 4343504           &4.8 \\
                               \hline
        \multirow{2}{*}{\textbf{12K}} & DTW          & $144 \times 10^{6}$           & N.A \\
                              & SparseDTW   & 7455538           & 200 \\                                                                                                                                                                                                                                   \hline\hline
    \end{tabular}
    \caption {\label{Tab:TSM-Results2} Performance of the \emph{DTW} and
    \emph{SparseDTW} algorithms using large datasets.}
\end{table}

\subsection{Discussion and Analysis}\label{Sec:TSM-DiscAnalysis}

\emph{SparseDTW} algorithm is evaluated against three other existing
algorithms, \emph{DTW}, which always gives the optimal answer,
\emph{DC}, and \emph{BandDTW}.

\begin{table*} 
    \centering
    \begin{tabular}{l l r r r r } \\ \hline \hline
        \multicolumn{1}{l}{Dataset}& \multicolumn{1}{l}{Algorithm}& \multicolumn{1}{l}{Number of }& \multicolumn{1}{l}{Warping path }&\multicolumn{1}{l}{Elapsed Time} &
        \multicolumn{1}{l}{DTW}
        \\
        name&name    & opened  cells   & size (K)&   (Seconds) &  Distance
        \\ \hline \hline
        \multirow{3}{*}{GunX} & DTW       & 22500         &201  & 0.016          & 0.01          \\
                              & BandDTW   & 448           &152  & 0.000          & 0.037         \\
                              & SparseDTW & 4804          &201  & 0.000          & 0.01          \\
                              \hline
        \multirow{3}{*}{Trace} & DTW      & 75076         &404  & 0.063          & 0.002         \\
                              & BandDTW   & 1364          &331  & 0.016          & 0.012         \\
                              & SparseDTW & 17220         &404  & 0.000          & 0.002         \\
                              \hline
       \multirow{3}{*}{Burst-Water}&DTW    & 2190000      &2190 & 1.578         & 0.102         \\
                               & BandDTW   & 43576        &2190 & 0.11          & 0.107         \\
                               & SparseDTW & 951150       &2190 & 0.75          & 0.102         \\
                               \hline
        \multirow{3}{*}{Sun-Spot} & DTW   & 1266610       &357 & 0.063            & 0.021         \\
                              & BandDTW   & 12457         &358 & 0.016            & 0.022         \\
                              & SparseDTW & 66049         &357 & 0.016            & 0.021         \\
                               \hline
        \multirow{3}{*}{ERP}  & DTW       & 1000000      &1533  & 0.78          & 0.008         \\
                              & BandDTW   & 19286        &1397  & 0.047         & 0.013         \\
                              & SparseDTW & 210633       &1535  & 0.18          & 0.008         \\                                                                                    \hline
        \multirow{3}{*}{Synthetic}  & DTW & 250000       &775  & 0.187          & 0.033          \\
                              & BandDTW   & 4670         &600  & 0.016         & 0.043           \\
                              & SparseDTW & 105701       &775  & 0.094          & 0.033          \\                                                                                                                                                                                                                                        \hline\hline
    \end{tabular}
    \caption {\label{Tab:TSM-Results} Statistics about the performance of \emph{DTW}, \emph{BandDTW}, and \emph{SparseDTW}.
    Results in this table represent the average over all queries.}

\end{table*}

\subsubsection{Elapsed Time}
The running time of the four approaches is shown in
Figure~\ref{Fig:TSM-ElapsedTime}. The time profile of both
\emph{DTW} and \emph{BandDTW} is similar and highlights the fact
that \emph{BandDTW} does not exploit the nature of the datasets.
\emph{DC} shows as well the worst performance due to the vast number
of recursive calls to generate and solve sub-problems.  In contrast,
it appears that \emph{SparseDTW} is exploiting the inherent
similarity in the GunX and Trace data.

 In Figure~\ref{Fig:TSM-ComputedCells} we show the number of
open/computed cells produced by the four algorithms. It is very
clear that \emph{SparseDTW} produces the lowest number of opened
cells.

In Table~\ref{Tab:TSM-ComputedCells-Diagonal} we show the number of
computed cells that are used in finding the optimal alignment for
three different datasets, where their optimal paths are close to the
diagonal. \emph{DC} has shown the highest number of computed cells
followed by \emph{DTW}. That is because both (\emph{DC} and
\emph{DTW}) do not exploit the similarity in the data.
\emph{BandDTW} has shown interesting results here because the
optimal alignment is close to the diagonal. However,
\emph{SparseDTW} still outperforms it.

Two conclusions are revealed from Figure~\ref{Fig:TSM-BandWidth}.
The first, the length of the time series affects the computing time,
because the longer the time series the bigger the matrix. Second,
band width influences CPU  time when aligning pairs of time series.
The wider the band the more cells are required to be opened.

\emph{DTW} and \emph{SparseDTW} are compared together using large
datasets. Table~\ref{Tab:TSM-Results2} shows that \emph{DTW} is not
applicable (N.A) for datasets of size $> 6K$, since it exceeds the
size of the memory when computing the warping matrix. In this
experiment we excluded \emph{BandDTW} and \emph{DC} given that they
provide no guarantee on the optimality.

To determine the effect of correlation on the elapsed time for
\emph{SparseDTW} we created several synthetic datasets with
different correlations. The intuition being that two sequences with
lower correlation will have a warping path which is further away
from the diagonal and thus will require more open cells in the
warping matrix. The results in Figure~\ref{Fig:TSM-Correlation}
confirm our intuition though only in the sense that extremely low
correlation sequences have a higher number of open cells than
extremely high correlation sequences.

\begin{figure*}[t]
  \begin{center}
    \mbox{
            \subfigure[GunX-DTW] {\includegraphics [width=0.32\textwidth]{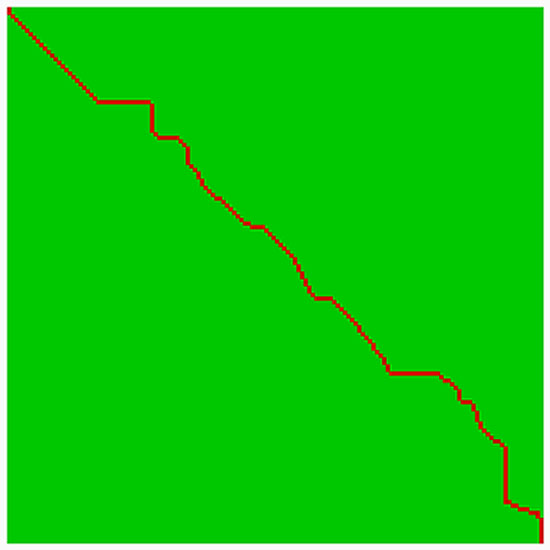}}
            \subfigure[GunX-BandDTW] {\includegraphics [width=0.32\textwidth]{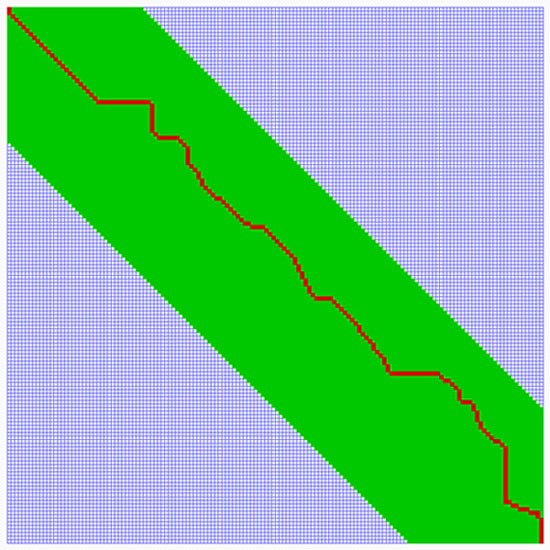}}
            \subfigure[GunX-SparseDTW] {\includegraphics [width=0.32\textwidth]{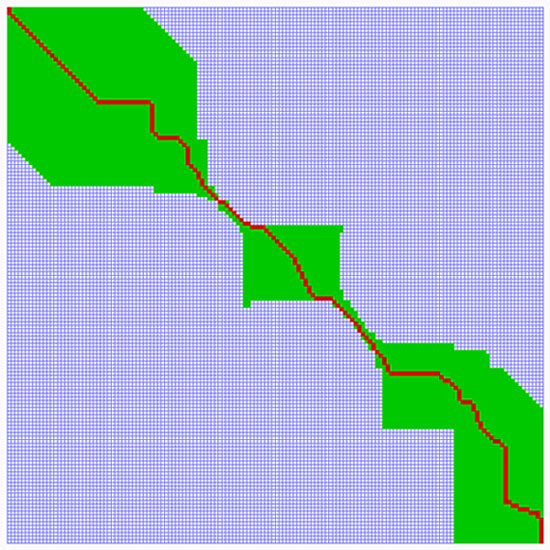}}
         }\\
    \mbox{
            \subfigure[Trace-DTW] {\includegraphics [width=0.32\textwidth]{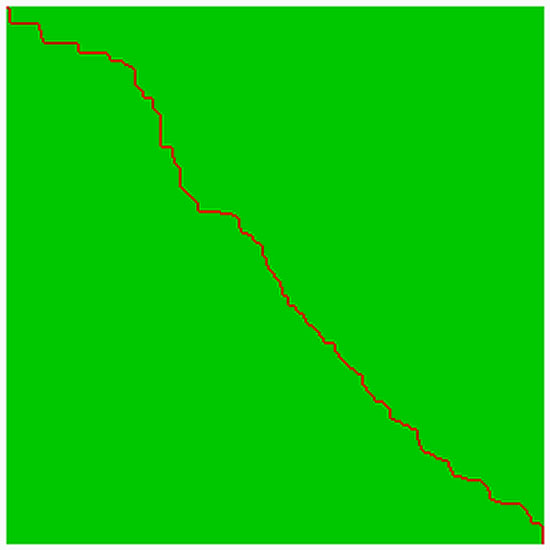}}
            \subfigure[Trace-BandDTW] {\includegraphics [width=0.32\textwidth]{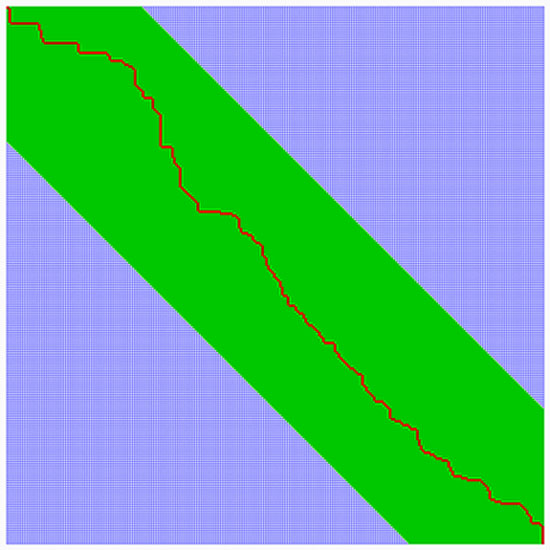}}
            \subfigure[Trace-SparseDTW] {\includegraphics [width=0.32\textwidth]{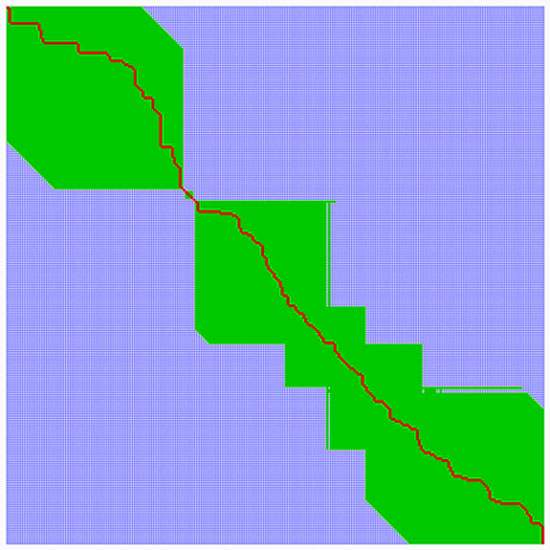}}
          }
      \caption{ The optimal warping path for the GunX and Trace sequences using three algorithms (\emph{DTW}, \emph{BandDTW}, and
               \emph{SparseDTW}). The advantages of \emph{SparseDTW} are clearly revealed as only a small fraction of the matrix cells have to be
               ``opened'' compared to the other two approaches.}
    \label{Fig:TSM-AllDTWAlgos}
  \end{center}
\end{figure*}

\subsubsection{SparseDTW Accuracy}

The accuracy of the warping path distance of \emph{BandDTW} and
\emph{SparseDTW} compared to standard \emph{DTW} (which always gives
the optimal result) is shown in Table~\ref{Tab:TSM-Results}. It is
clear that the error rate of \emph{BandDTW} varies from 30\% to
500\% while \emph{SparseDTW} always gives the exact value.
 It should be noticed that there may be more than one
optimal path of different sizes but they should give the same
minimum cost (distance). For example, the size of the warping path
for the \emph{ERP} dataset produced by \emph{DTW} is 1533, however,
\emph{SparseDTW} finds another path of size 1535 with the same
distance as \emph{DTW}.

Figure~\ref{Fig:TSM-AllDTWAlgos} shows the dramatic nature in which
\emph{SparseDTW} exploits the similarity inherent in the sequences
and creates an adaptive band around the warping path. For both the
GunX and the Trace data, \emph{SparseDTW} only opens a fraction of
the cells compared to both standard \emph{DTW} and \emph{BandDTW}.


\section{Conclusions} \label{Sec:PT-Summary}

In this paper we have introduced the \emph{SparseDTW} algorithm,
which is a sparse dynamic programming technique. It exploits the
correlation between any two time series to find the optimal warping
path between them. The algorithm finds the optimal path efficiently
and accurately. \emph{SparseDTW} always outperforms the algorithms
\emph{DTW}, \emph{BandDTW} and \emph{DC}. We have shown the
efficiency of the proposed algorithm through comprehensive
experiments using synthetic and real life datasets.

\bibliographystyle{agsm}
\bibliography{ThesisBib}

\end{document}